# Topological phase transition of the centered rectangular photonic lattice

J. Hajivandi, H. Kurt

Nanophotonics Research Laboratory, Department of Electrical and Electronics Engineering, TOBB University of Economics and Technology, Ankara 06560, Turkey

*jamileh@etu.edu.tr

**Abstract**

A Cm planar photonic material (two-dimensional) including the mirror reflection symmetry is explored where in the Dirac cones appeared at the high-symmetry points of the Brillouin zone boundary. By implementing the specific perturbation on the photonic crystal (PC), topological transition can make a bridge between the trivial PC (ordinary insulator) and the topological insulator including zero and non-zero spin Chern number (Cs), respectively. The perturbation can be realized through the rotation of the Cm photonic crystal around the nonzero angle to its initial position. Therefore, breaking the mirror symmetry of the unit cell leads to mismatch of the symmetry planes of the lattice. This modification results in the directional band gap & band inversion which is the signature of the topological transitions. The mentioned PC would be suitable for examination of the unidirectional transport of light at the topological interfaces.

## 1- Introduction

The enormous applications of the conception of topology in electronics, phononics and photonics, make it fascinating for studying the robust one-way transport of the edge states against different defects in topological insulators (TIs) [1-8].
Then, studying the topological characterization of the photonic structures is significant from scientific point of view. In the recent work, the photonic topological edge modes are realized equivalent to the quantum spin Hall effect (QSHE) [9]. In addition, Time reversal (TR) symmetry breaking, then unidirectional edge waves are proposed by temporal modulation of PCs which is the principal of the Floquet TIs [10] or the photonic Floquet TIs are realized in the absence of the external fields through an array of the helical waveguides coupled evanescently in a honeycomb lattice to exhibit the topologically protected edge states by preserving the TR symmetry [11].
The PCs with $C_6$ or $C_4$ point group symmetries, possess a four-fold degeneracy (double Dirac cones) [12-18] or three-fold degeneracy (Dirac-like cone) at the Brillouin zone center [19-21]. These kind of degeneracies are so sensitive to the lattice geometrical modifications. The degenerated bands at the Dirac cone can be inversed if the lattice experiences the inversion of symmetry, then the topological phase transition will be observed [22-30].
Moreover, a two-fold degeneracy (single Dirac cone) at the $K$ point of the Brillouin zone can be appeared in the honeycomb lattices. Then, the valley spin Hall effect (VSHE) can be explored by breaking the mirror reflection symmetry of the lattice arising from the mismatch of the symmetry planes of the lattice. This kind of Dirac cone is not sensitive to the geometric changes of the structure [31-40].
In our recent works, slow light and robust transport of topological edge states have been explored through the QSHE by implementing the honeycomb lattice including six air holes on the core-shell medium [41]. Likewise, preserving the Fano resonance mode and directional surface edge modes have been realized in the honeycomb PCs where in the $C_6$ point group symmetry is reduced to the $C_3$ one by breaking the mirror reflection symmetry of the PCs [42-44].
Here, the unavoidable band degeneracy of a centered rectangular, Cm planar lattice has been investigated which leads to appearing the two single Dirac cones at the points $Y$ & $Y_1$. By breaking the mirror reflection symmetry of the PC, the Dirac cones will be opened. Applying rotations from right to left or left to right induces the inversed mirror reflection symmetries. This closing/opening the Dirac cones and band inversion at the points $Y$ & $Y_1$, similar to the valley-like coupling, convince the topological phase transition. The emergence and propagation of the edge states at the topological straight interface has been studied numerically in this paper.

## 2- Centered rectangular PC with the Cm group symmetry

The band degeneracy of the centered rectangular PC including the Cm point group symmetry has been explored in this work. Due to the closing of the band gap and then expected the band degeneracy, two single Dirac cones are appeared in the PC. Breaking the mirror-reflection symmetry leads to opening of the bandgaps near the Dirac points. By applying opposite rotations from right to left or left to right, the inversed mirror reflection symmetries have been observed. The closing/opening the Dirac cones and band inversion at the Dirac cones is a signature of a topological transition. Here, transport of the topological edge states in the Cm lattice, has been demonstrated numerically.

Indeed, Cm is including the letters "C" and "m" which indicate the centered rectangular lattice with the reflection symmetry (mirror /line), respectively. The group Cm have one mirror reflection symmetry plane, $m_{d1}$ [45]. The primitive unit cell and Bravais lattice of the Cm planar lattice including the mirror reflection symmetry, $m_{d1}$ and the Brillouin zone are presented in the Fig. 1. The lattice vectors are $\vec{a} = (a, 0)$ and $\vec{b} = (0, b)$ with the lattice constant $a$ and $b$. Transforming the basic unit cell in the direction of the lattice vectors $\vec{a}$ and $\vec{b}$, lead to generating the complete Bravais lattice.

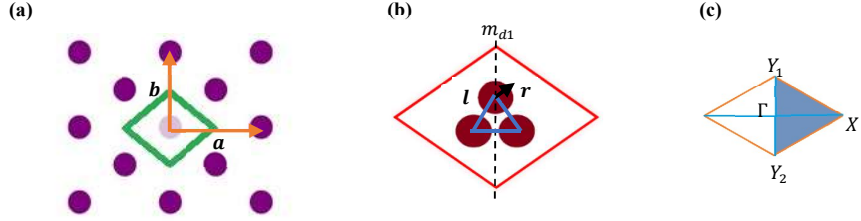

Fig. 1. (a) The Bravais Lattice, (b) the Primitive unit cell with mirror symmetry, $m_{d1}$ and (c) the Brillouin zone of the Cm planar PC.

Each unit cell of the Cm planar PC consists of three circular rods with the radius $r$ and the dielectric $\varepsilon = 12$ surrounded by the air. The distances between centers of rods are identical and equals to $l$. For dispersion calculation, the plane-wave expansion scheme by means of the package *MIT photonic bands (MPB)* has been employed [46]. Fig. 2. presents the bandstructure of the Cm photonic configuration for the TM modes.
In continue, one may see the Dirac cones $N$ ($N'$) and the degeneracy frequencies at $Y_1$ ($Y_2$) points as shown in the Fig. 2. for the parameters $a = 1$, $b = 0.8$, $l = 0.45\,a$ and $r = 0.5l$.

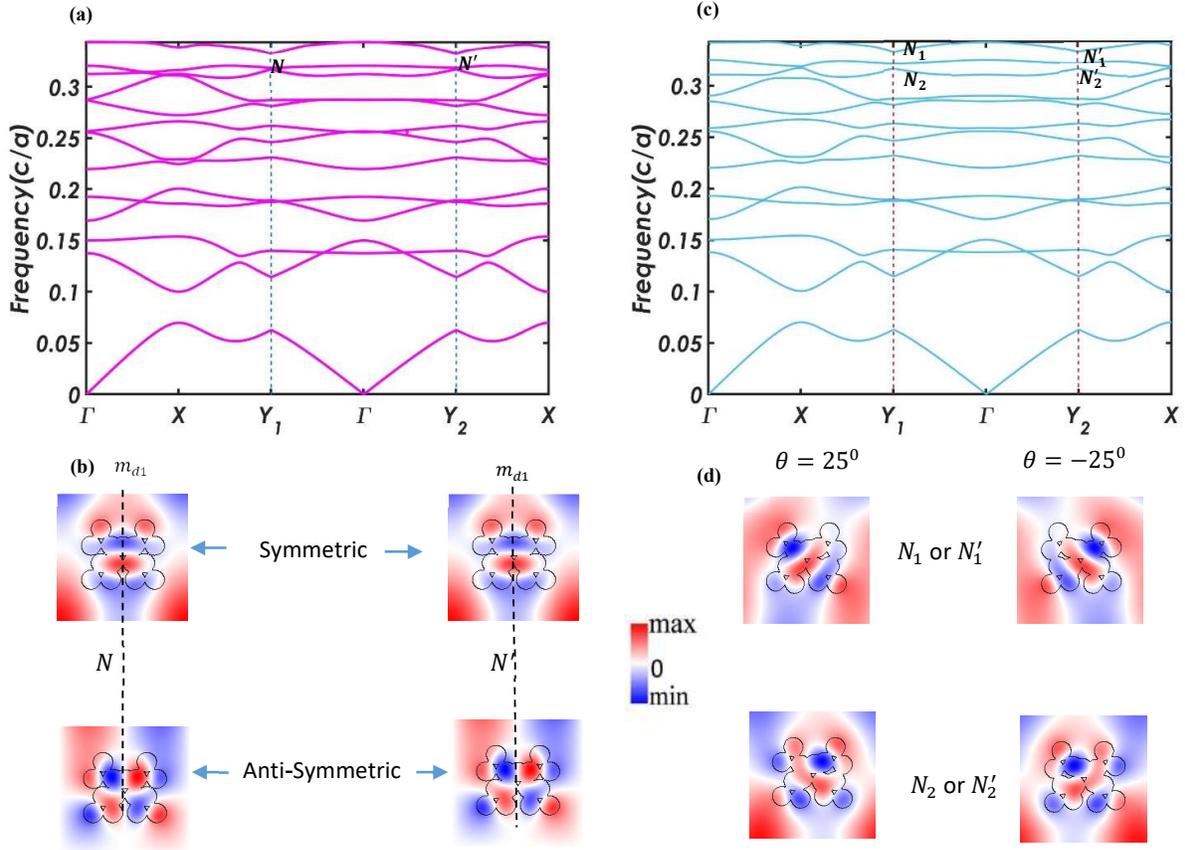

Fig. 2. (a) Bandstructure graph for the Cm planar photonic structure for: $a = 1$, $b = 0.8$, $l = 0.45\,a$ and $r = 0.5l$, (b) The identical symmetric and antisymmetric eigenmode profiles, $E_z$ of the Dirac cones $N$ (left) & $N'$(right), (c) The dispersion diagram of the Cm lattice by rotating around the $\theta = 25^0$, (d) Inversed eigenmodes, $E_z$ of points $N_1, N_2$ ( or $N_1', N_2'$) for angels $\theta = 25^0$ and $\theta = -25^0$.

As seen from the Fig. 2. the Dirac points $N$ and $N'$ indicate identical but symmetric/anti-symmetric eigenmodes, $E_z$, about the mirror axis $m_{d1}$. By rotating this Cm planar PC columns at angle $\theta = 25^0$ relative to their initial position, the Dirac cones $N$ and $N'$ will be opened as points $N_1$, $N_2$ and $N'_1$, $N'_2$, so they become nondegenerate. In fact, the mismatch between the axis of the rods and the crystal, due to the rotation of the PC around the specific $\theta$, leads to the mirror symmetry breaking and opening of the Dirac points, see Fig. 2 (c). We note that there is the same diagram as Fig. 2 (c) around the angle of the rotation $\theta = -25^0$. In Fig. 2 (d) the eigenmode profiles of the points $N_1$, $N_2$ ( or $N'_1$, $N'_2$ ) for the rotation around the angels $\theta = 25^0$ and $\theta = -25^0$, are shown where in the phases of the eigenmodes at $\theta = -25^0$ are inversed from the phases of eigenmodes at $\theta = 25^0$.

Additionally, in Fig. 3., we study the angular dependent frequencies of points $N_1$, $N_2$ ( or $N'_1$, $N'_2$ ), for the parameter $a = 1$, $b = 0.95$, $l = 0.53\ a$ and $r = 0.1l$, where in the opening, closing and reopening of the Dirac points through the rotations $-\theta \to 0 \to \theta$ can be observed. Moreover, we found that the bands have inverted at $\theta = 0$, the red and green bands indicate the frequency of the points $N_1$ ($N'_1$) and $N_2$ ($N'_2$) respectively. Thus, through the exchanging of the degenerated bands and inversing the eignstates at $\theta = 0$, topological phase transitions will be induced.

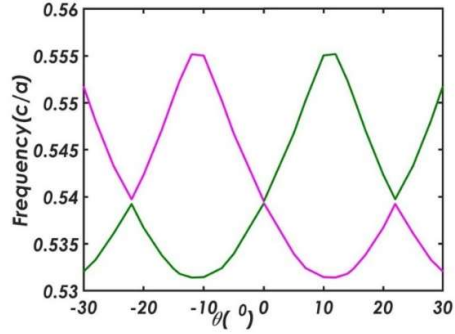

Fig. 3. Phase diagram of the Cm planar PC versus angle $\theta$ for the parameters $a = 1$, $b = 0.95$, $l = 0.53\ a$ and $r = 0.1l$, the red and green bands indicate the frequency of the points $N_1$ ($N'_1$) and $N_2$ ($N'_2$) respectively

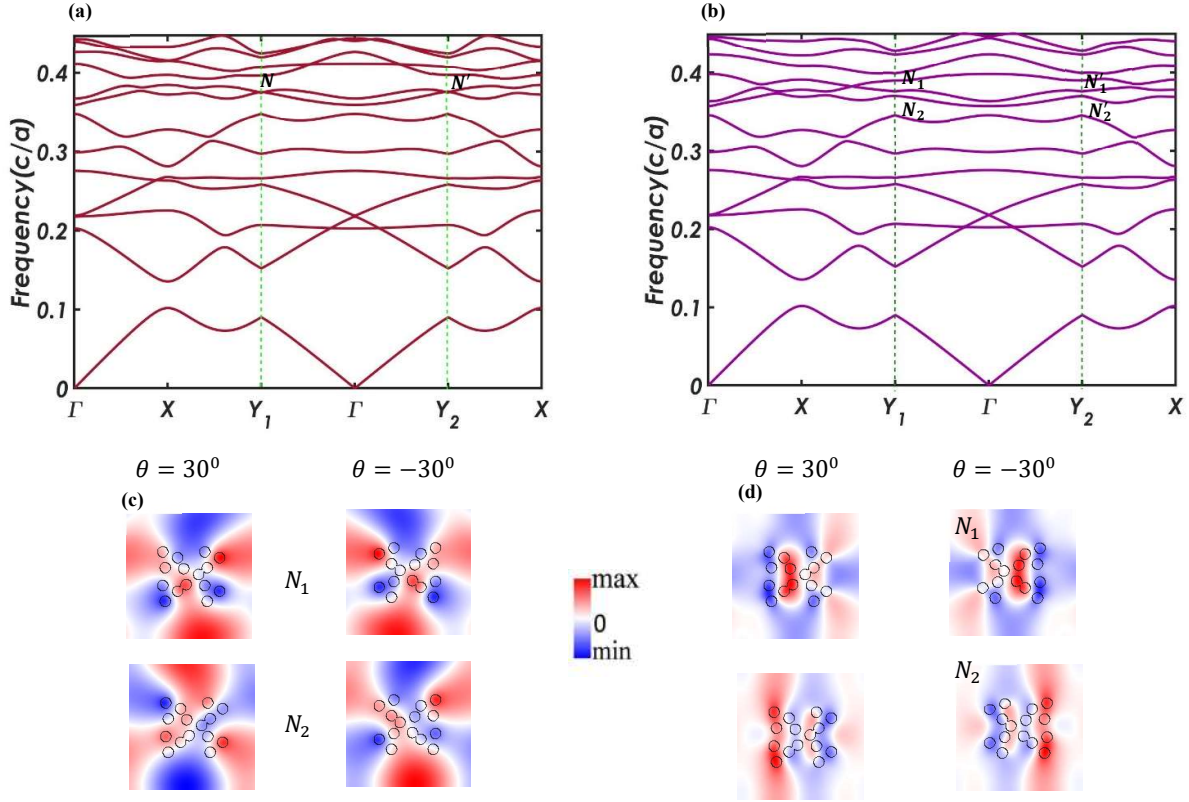

Fig. 4. The bandstructure graph of the Cm planar lattice with $a = 1$, $b = 0.8$, $l = 0.38\ a$ and $r = l/3$ for TM polarization and the rotation at (a) $\theta = 0$, (b) $\theta = -30^0, -30^0$, (c) The field profiles of the points $N_1$, $N_2$ and (d) the pointing vector profiles of them.

## 3- Emergence of the topological edge modes in the centered rectangular lattice

For studying the one-way propagation of the edge states at the topological interfaces, we can design the upper/downer photonic topological insulator composed of two types of PCs including common bandgap but opposite band topology. As found from the previous section, by rotating the Cm PC around the angles $\theta$ and $-\theta$, therefore opening, closing and reopening of the Dirac points, the topological phase transition will be understand. In continue, we design an photonic topological insulator (PTI) with the upper/downer supercell which is composed of 10 (10) PCs with rotations $\theta = -30^0$ ($\theta = 30^0$) respectively and for the parameters $a = 1$, $b = 0.8$, $l = 0.38\,a$ and $r = l/3$. We will study the edge state behavior of the PTI for TM polarization. Fig. 4. shows the band diagram, field profiles and pointing vectors of the points $N_1$, $N_2$ for the rotations at $\theta = -30^0$ and $\theta = 30^0$.

As indicated in Fig. 4. (c)-(d), the field profiles, $E_z$ of points $N_1$, $N_2$ ( or $N_1'$, $N_2'$ ) and their pointing vectors can be inverted to each other by rotation $\theta = 30^0$ to $\theta = -30^0$.

By implementing two PCs with same bandgap but different topology for designing an upper/downer PTI, the emergence and transport of the edge states can be studied. Fig. 4. Shows the band structure of the PTI and the edge modes, A & B, their field and pointing vector profiles.

The *Lumerical FDTD* simulation has been performed to study the robust topological edge states for the upper/downer design system is bounded through the Perfectly Matching Layers (PMLs) [47].

Fig. 5. shows the one-way transport of the edge states without noticeable back-scattering at the simulation region. To produce the out of plane electric field $E_z$ along the interface $+x$, two in-plane magnetic dipole sources with the phase difference $90^0$ between them ($H_x + iH_y$), has been applied at the point indicated with the yellow star in the Fig. 5. The excitation frequency is $0.343371$ ($^c/_a$).

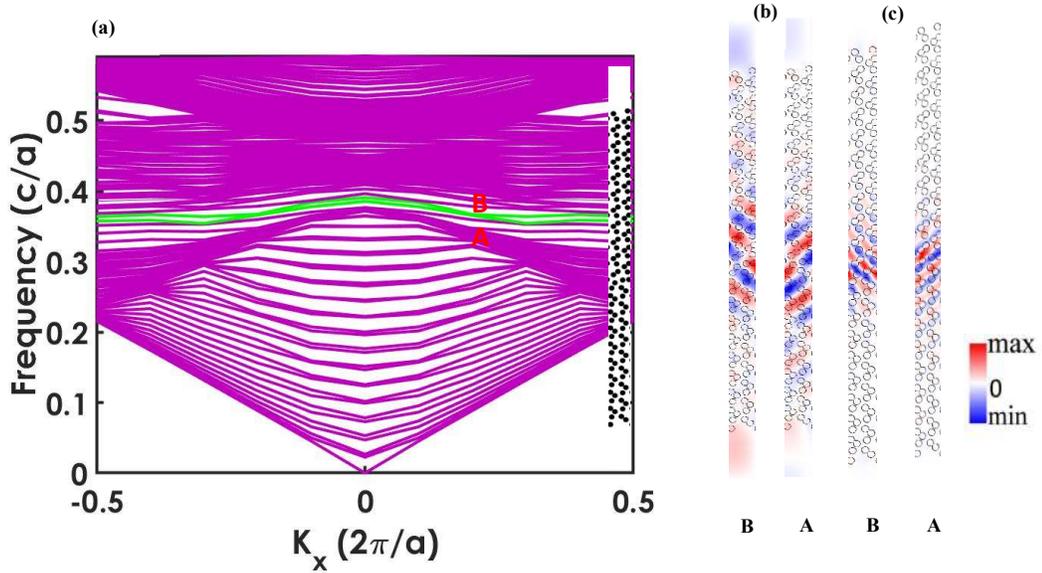

Fig. 4. (a) The band structure of a PTI with the upper/downer supercell which is composed of 10 (10) PCs with rotations $\theta = -30^0$ ($\theta = 30^0$) respectively and for the parameters $a = 1$, $b = 0.8$, $l = 0.38\,a$ and $r = l/3$ with the edge modes, A & B, (b) the $E_z$ sketches of the edge modes A & B and (c) pointing vector profiles of these the edge modes.

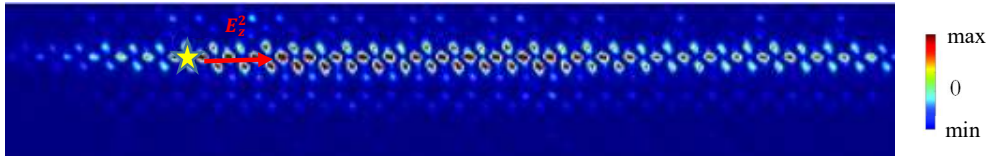

Fig.5. The one-way transport of the edge modes at the topological interface of the Cm planar lattice.

## 4- Conclusion

In this paper, we have investigated the Dirac cones of a Cm planar PC along the high-symmetry line in the Brillouin Zone. Besides, we found that breaking of the mirror reflection symmetry lead to removing of the Dirac cones. So an opening-closing and reopening of the Dirac cones have been observed through the rotation around $-\theta \to 0 \to \theta$ which is the signature of the topological transitions.

Moreover by studying the phase diagram of the structure along the various angles, we found that the bands have inverted at $\theta = 0$. Also the phases of the eigenmodes at $\theta$ are inversed from the phases at $-\theta$. Thus, the topological phase transition can be induced through the band switching and eigenmode inverting at $\theta = 0$. By designing the upp/down PTI, we found the robust transport of the topological edge modes at the straight interface.

## Acknowledgment

H. K. acknowledges partial support of the Turkish Academy of Sciences.